\documentclass[runningheads]{llncs}
\usepackage{graphicx}
\usepackage{amssymb}
\usepackage{svg}
\usepackage{amsmath}
\usepackage{comment}
\usepackage{tabularx}
\usepackage{mathtools}
\usepackage{subcaption}
\usepackage[english]{babel}
\usepackage{wrapfig}
\usepackage{multirow}
\newcommand{\mycomment}[1]{%
}%

\usepackage{listings}
\usepackage{color}

\definecolor{dkgreen}{rgb}{0,0.6,0}
\definecolor{gray}{rgb}{0.5,0.5,0.5}
\definecolor{mauve}{rgb}{0.58,0,0.82}

\begin{document}
\title{Learning Positional Attention for Sequential Recommendation}
%
%
\author{Fan Luo\and
Haibo He\and
Juan Zhang\and
Shenghui Xu
}
\institute{NetEase Media Technology (Beijing) Co., Ltd.}

\maketitle              
\begin{abstract}
Self-attention-based networks have achieved remarkable performance in sequential recommendation tasks. A crucial component of these models is positional encoding. In this study, we delve into the learned positional embedding, demonstrating that it often captures the distance between tokens. Building on this insight, we introduce novel attention models that directly learn positional relations. Extensive experiments reveal that our proposed models, \textbf{PARec} and \textbf{FPARec} outperform previous self-attention-based approaches. The code can be found here: https://github.com/NetEase-Media/FPARec. 

\end{abstract}
%
%
%
\section{Introduction}   
Recommendation systems play pivotal roles in the era of information explosion, finding widespread application in e-commerce and online media platforms, such as Amazon and YouTube. These systems predict items users might be interested in, effectively mitigating the challenge of information overload. Sequential recommendation, in particular, aims to predict user interests based on their past behavior sequences.

In 2017, the Transformer architecture \cite{transformer} was introduced and rapidly demonstrated superior performance in sequence modeling tasks. Subsequently, it was adapted for sequential recommendation tasks, giving rise to models like SAS4Rec \cite{sasrec} and BERT4Rec \cite{BERT4Rec}. Central to both is the dot-product self-attention mechanism, which allows any token to attend to all other tokens, facilitating the learning of long-term dependencies.

Yet, a limitation of self-attention is its insensitivity to position, rendering it incapable of accounting for token order. To address this, positional embeddings were introduced to capture the sequential order of tokens. These embeddings, which can be learned alongside other model parameters, raise an intriguing question: how does the learned positional embedding influence the attention mechanisms? Based on empirical analysis, we posit that in sequential recommendation tasks, positional embeddings predominantly capture distances between tokens. 

Motivated by this insight, we propose two novel models, \textbf{P}ositional \textbf{A}ttention network for sequential \textbf{Rec}ommendation \textbf{(PARec)} and \textbf{F}actorized \textbf{P}ositional \textbf{A}ttention network for sequential \textbf{Rec}ommendation \textbf{(FPARec)}. \textbf{PARec} introduces novel positional attention mechanism that employ a weight matrix to learn attention patterns between different positions.   \textbf{FPARec} factorizes this attention matrix, reducing the parameter count of the attention layer. To the best of our knowledge, it is the first time that such approach is used for sequential recommendation, resulting in simple and effective models. Through extensive experiments across multiple real-world datasets, our methods are shown to surpass state-of-the-art self-attention-based sequential recommendation models in performance.

\section{Preliminary}
In this section, we review the self-attention mechanism for sequential recommendation model and discuss the effect of positional encoding on attention layers.

\subsection{Self-attention Mechanism for Sequential Recommendation Model}

In sequential recommendation tasks, we are given the user's interaction sequence \(\mathcal{S}_u = [v_1^u, \ldots, v_t^u, \ldots, v_{n_u}^u ]\) in chronological order. Here, \(v_t^u\) is the \(t\)-th item that user \(u\) interacted with, and \(n_u\) represents the total length of interaction history for user \(u\). The task aims to predict the next item that user \(u\) will interact with.

Given this framework of sequential recommendation tasks, self-attention model is utilized capture the dependencies and patterns in user interaction sequences. In self-attention model using scale dot-product attention mechanism, the output of self-attention layer for position \(i\) is computed as 

\begin{equation}
Attention(\mathbf{Q}, \mathbf{K}, \mathbf{V})_i = \frac{\sum_{j=1}^n \text{sim}(Q_i,K_j)V_j} {\sum_{j=1}^n\text{sim}(Q_i,K_j)}
\end{equation}

\begin{equation}
\text{sim}(Q_i,K_j) = \exp\left(\frac{Q_iK_j^T}{\sqrt{d}}\right) \label{eq:sim}
\end{equation}

Here, \(Q_i\) and \(K_j\) are query and key respectively. \(V_j\) is value for position \(j\). \(d\) is the number of hidden dimension of the input embedding.

\subsection{Positional Embedding Analysis}

Much of the previous research on positional embedding has been centered around models trained on natural language corpora\cite{RethinkingPE,WhatDP,OnPE}. The correlation between positional embeddings corresponding to different positions has typically been explored by calculating the dot-product or cosine similarity between positions. Representing positional embedding for query and key as \(P_i\) and \(P_j\) respectively, we propose that it's more meaningful to use \(\exp{(\frac{P_iP_j^T}{\sqrt{d}})}\) to gauge the correlation between positions since the \(exp(\bullet)\) operation is integral to the dot-product attention mechanism.

To elucidate this, recall that in the first self-attention layer, the similarity between the query and key is a composite of four terms:
\begin{equation}
\begin{split}
sim(Q_i, K_j) 
= & \underbrace{\exp{(\frac{M_i\bold{W_Q}(M_j\bold{W_K})^T}{\sqrt{d}})}}_\text{item-item attention}
\underbrace{\exp{(\frac{M_i\bold{W_Q}(P_j\bold{W_K})^T}{\sqrt{d}})}}_\text{item-position attention} \\
& \underbrace{\exp{(\frac{P_i\bold{W_Q}(M_j\bold{W_K})^T}{\sqrt{d}})}}_\text{item-position attention}
\underbrace{\exp{(\frac{P_i\bold{W_Q}(P_j\bold{W_K})^T}{\sqrt{d}})}}_\text{position-position attention}
\end{split}
\end{equation}

\begin{equation}
sim(Q_i, K_j) \propto \exp{(\frac{P_i\bold{W_Q}(P_j\bold{W_K})^T}{\sqrt{d}})}
\end{equation}

Here, \(M_i\) and \(M_j\) are item embeddings for query item and key item respectively. \(\mathbf{W_Q}\) and \(\mathbf{W_K}\) are weight matrices for query projection and key projection.

\begin{figure}[b]
    \vspace{-10pt}
    \centering
    \begin{minipage}{0.45\textwidth}
        \centering
        \includegraphics[width=0.5\textwidth]{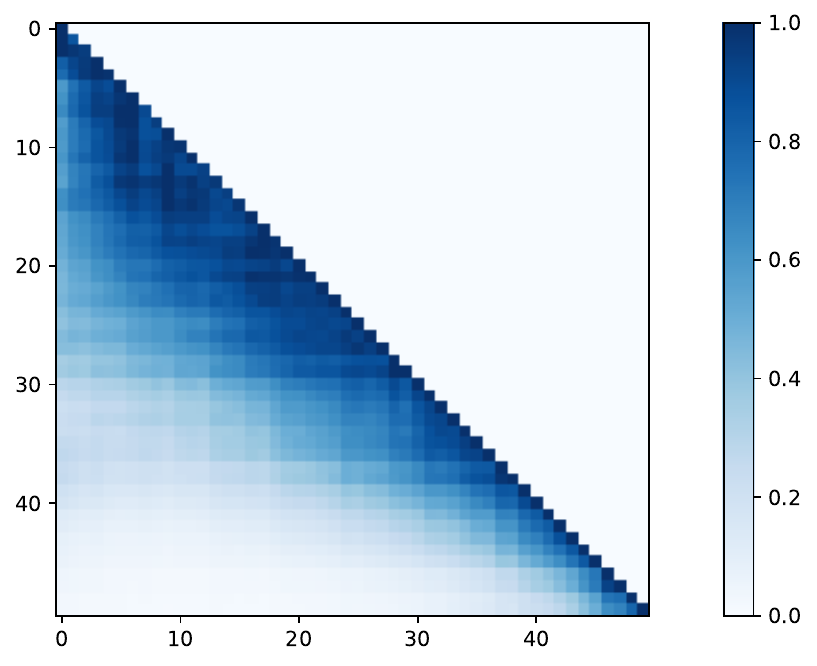} 
        \subcaption{maximum input sequence length = 50}
    \end{minipage}\hfill
    \begin{minipage}{0.45\textwidth}
        \centering
        \includegraphics[width=0.5\textwidth]{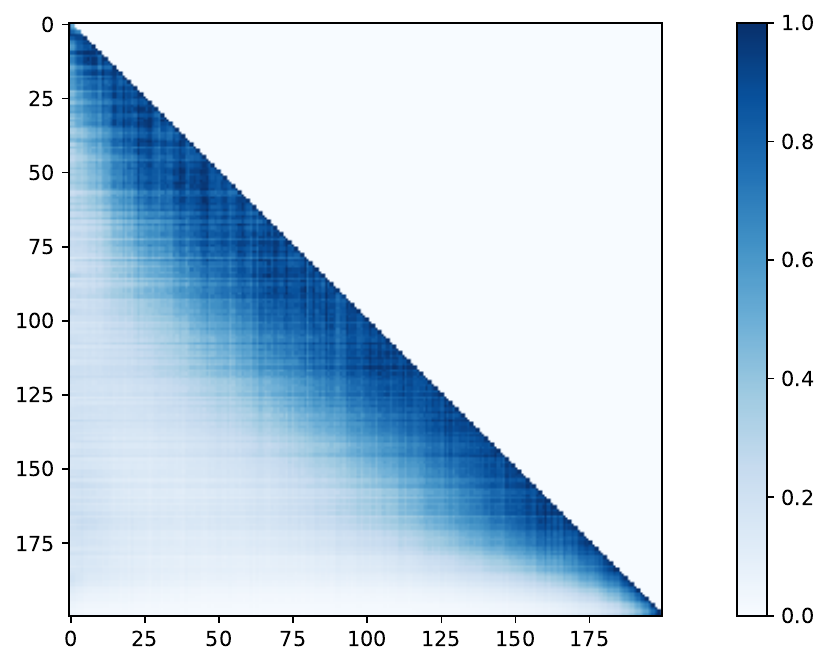} 
        \subcaption{maximum input sequence length = 200}
    \end{minipage}
    \caption{Correlation between position embeddings, normalized by the maximum value in each respective row for clarity in visualization. Further details on producing this figure are available in the supplementary materials.}  \label{fig:correlation}
    \vspace{-1em}
\end{figure}

It's evident that the actual attention is a product of these four components. The last term accounts for the interaction between positions. The magnitude of this interaction is proportional to \(e\) raised to the power of the dot-product of the linearly projected positional embedding. Given that some models might employ multiple heads in a single self-attention layer—with each head potentially using different projection matrices—we chose to solely examine \(\exp{(\frac{P_iP_j^T}{\sqrt{d}})}\). Moreover, since the model being investigated utilizes unidirectional attention, we zero out the correlation values associated with future positions. The derived correlations are depicted in figure \ref{fig:correlation}.

Figure \ref{fig:correlation} suggests that positions closer together have a higher correlation than those farther apart. This implies that the positional embedding assigns varying importance to different positions, giving more emphasis to adjacent positions. In light of this understanding, we will unveil a new methodology that harnesses this characteristic in the next section.

\section{Method}


In this section, we describe our models, \textbf{PARec} and \textbf{FPARec} which utilize a novel attention mechanism that can automatically learn suitable positional attention patterns for various sequential recommendation scenarios.

\subsection{Model Architecture}

In this subsection, we illustrate our model in detail. Our model consists of an embedding layer, multiple stacked positional attention blocks, and a prediction layer. 

\begin{figure}[t]
    \vspace{-1em}
    \centering
    \begin{minipage}{0.3\textwidth}
        \centering
        \includegraphics[width=0.9\textwidth]{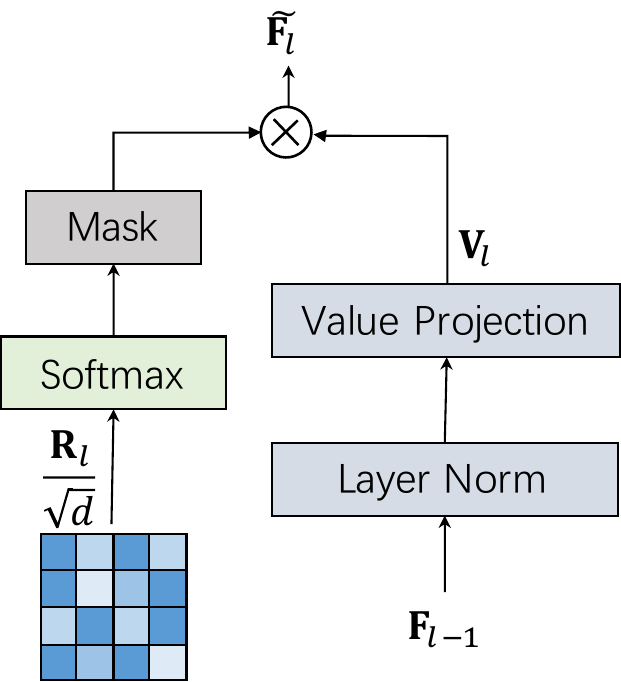} 
        \subcaption{}
    \end{minipage}\hfill
    \begin{minipage}{0.3\textwidth}
        \centering
        \includegraphics[width=0.9\textwidth]{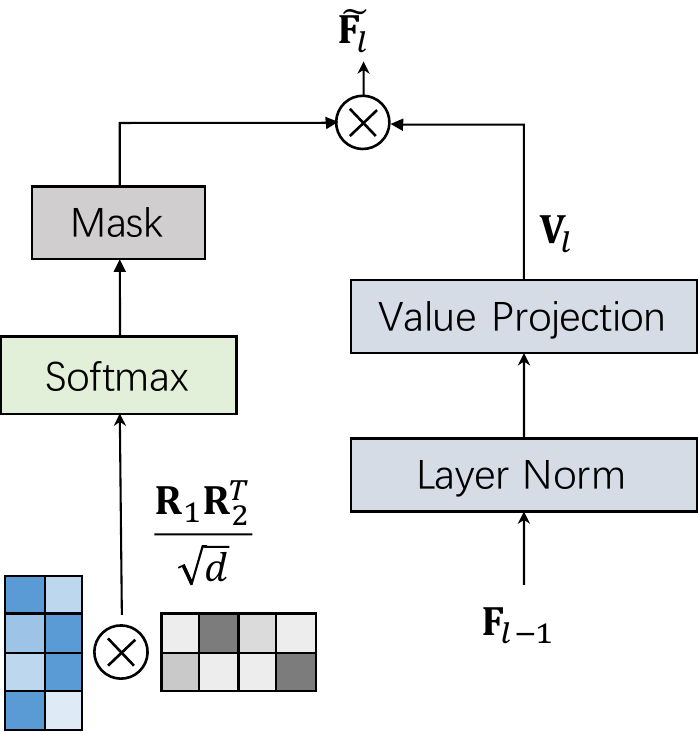} 
        \subcaption{}
    \end{minipage}\hfill
    \begin{minipage}{0.35\textwidth}
        \centering
        \includegraphics[width=0.9\textwidth]{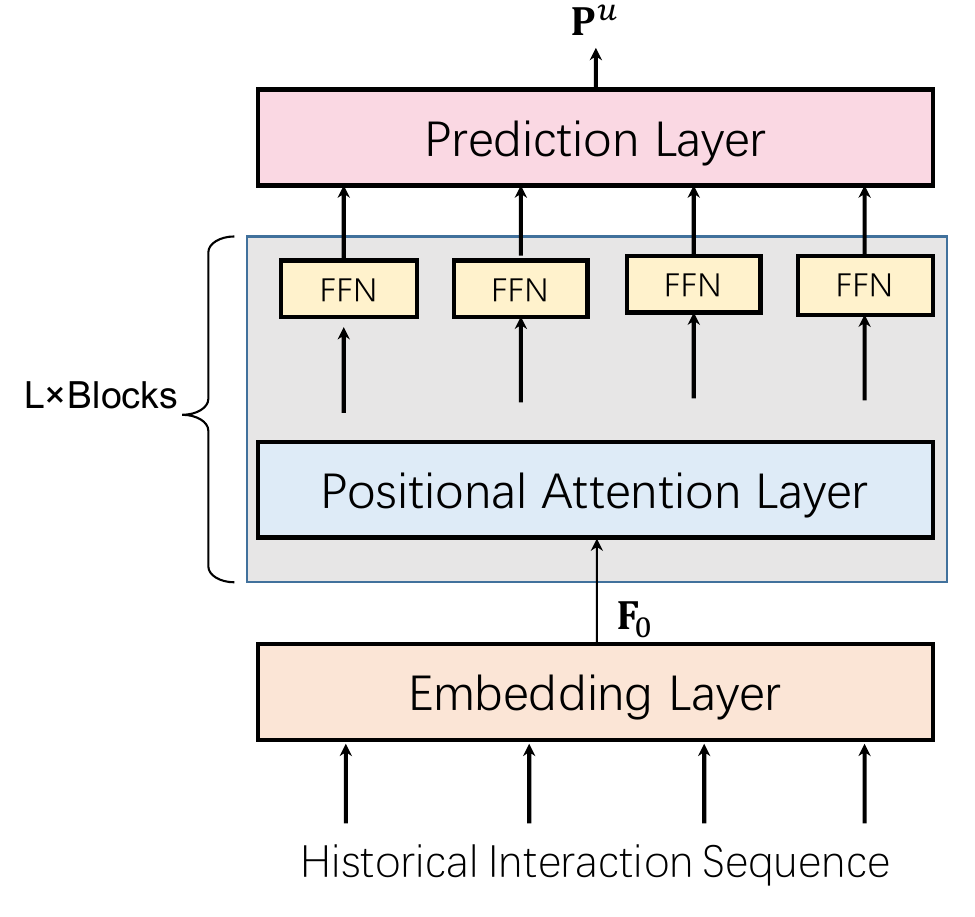} 
        \subcaption{}
    \end{minipage}
    \vspace{-1em}
    \caption{Model architecture. (a) Positional attention layer using a single learnable matrix \(\mathbf{R}_l\) to capture positional relation. (b) The variant of positional attention layer using two small matrices \(\mathbf{R}_{1,l}\) and \(\mathbf{R}_{2,l}\) to capture positional relation. (c) The overall architecture of our models. }
    \vspace{-1em}
\end{figure}
\subsubsection{Embedding Layer}

In the embedding layer, a user's historical interaction sequence is fed into the layer. A embedding matrix \(\mathbf{M}\in \mathbb{R}^{|\mathcal{V}|\times d}\) is created to map the item ids into \(d\)-dimensional vectors. The lookup result is denoted as \(\mathbf{F}_0\), and \(\mathbf{F}_0\in\mathbb{R}^{n\times d}\).

\subsubsection{Positional Attention Block}
We stack multiple Positional attention blocks on top of the embedding layer. Each positional attention block consists of a positional attention layer and a 2-layer position-wise feed-forward network.

\paragraph{Positional Attention Layer}: 
Let \(\mathbf{F}_{l-1}\) denote the input to \(l\)-th positional attention layers, the feature from the previous layer is normalized using layer normalization \cite{layer_normalization} and then linear projected into the value matrix \(\mathbf{V}_l\):

\begin{equation}
\mathbf{V}_l = LayerNorm(\mathbf{F}_{l-1})\mathbf{W}_{V, l}
\end{equation}

Here, \(\mathbf{W}_{V, l}\) represents the weight matrix for value projection in the \(l\)-th layer, which is shared across all positions within the same layer. Upon obtaining the value matrix, a learnable matrix \(\mathbf{R}_l\) is then applied to compute the output of the positional attention layer, integrating the learned positional relationships:

\begin{equation}
\mathbf{\tilde{F}}_l = Mask(softmax(\frac{\mathbf{R}_l}{\sqrt{d}}))\mathbf{V}_l \label{eq:pa}
\end{equation}

The learnable matrix \(\mathbf{R}_l\), unique to each \(l\)-th positional attention layer, captures the positional relations within the sequence. It is scaled by \(\frac{1}{\sqrt{d}}\) and then normalized using the \(softmax\) function to ensure that attention weights remain positive. To maintain the autoregressive property of the model, we apply a masking operation that uses a lower triangular matrix, preventing the layer from attending to future positions in the sequence. The model variant incorporating this attention mechanism is designated as \textbf{PARec}. 

To optimize parameter efficiency, the matrix \(\mathbf{R}_l\) is factorized into two lower-rank matrices \(\mathbf{R}_{1,l}\) and \(\mathbf{R}_{2,l}\), both of dimension \(\mathbb{R}^{n\times k}\). These matrices approximate the original \(\mathbf{R}_l\) through their product, thereby capturing positional interactions with fewer parameters:

\begin{equation}
\mathbf{\tilde{F}}_l = Mask(softmax(\frac{\mathbf{R}_{1,l}\mathbf{R}_{2,l}^T}{\sqrt{d}}))\mathbf{V}_l \label{eq:pa}
\end{equation}

This approach leverages the factorized positional attention to reduce the model's complexity while maintaining the ability to learn rich positional patterns. We refer to the model using factorized positional attention as \textbf{FPARec}.

\paragraph{Position-Wise Feed-Forward Network}: In addition to the positional attention layer, We append a feed-forward network to the attention block. Representation in each position passes the network separately,  

\begin{equation}
\mathbf{F}_l = FFN(LayerNorm(\mathbf{\tilde{F}}_l)) =  ReLU(LayerNorm(\mathbf{\tilde{F}}_l)\mathbf{W}_{1, l} + \mathbf{b}_{1, l})\mathbf{W}_{2, l} + \mathbf{b}_{2, l}
\end{equation}

\(\mathbf{W}_{1, l}\), \(\mathbf{W}_{2, l}\), \(\mathbf{b}_{1, l}\), \(\mathbf{b}_{2, l}\) are weights and biases for feed-forward network in \(l\)-th attention block. 

\subsubsection{Prediction Layer}

We apply layer normalization on the output of the last attention blocks \(\mathbf{F}_L\) and multiply it by the weight of the prediction layer,

\begin{equation}
\hat{\mathbf{F}}_L = LayerNorm(\mathbf{F}_L)
\end{equation}
\begin{equation}
\mathbf{P}^u = \hat{\mathbf{F}}_L\mathbf{M}^T
\end{equation}

\(\mathbf{P}^u \in \mathbb{R}^{n \times |\mathcal{V}|} \) is the prediction for user \(u\). The predicted user preference at time step \(t\) for item \(v\) can be seen as the dot product between \((t-1)\)-th row of \(\hat{\mathbf{F}}_L\) and embedding of that item \(M_v\) and is denoted as \(P^u_{t-1,v}\) for later use.  

\subsubsection{Model Training}

We employ cross-entropy loss on all items to train our model. 
Let \(P^u_{t-1,v}\) denote the predicted preference score for ground-truth item \(v_t^u\) of user \(u\) at time step \(t\). The loss is defined as the negative log-likelihood of that ground-truth item: 

\begin{equation}
\mathcal{L}=\frac{1}
{\sum_{u \in \mathcal{U}}|\mathcal{S}^u|}
\sum_{u\in \mathcal{U}} \sum_{t=n-|\mathcal{S}_u|+1}^{n+1} 
-log \frac{\exp P^u_{t-1, v^u_t}}{\sum_{v\in\mathcal{V}}\exp P^u_{t-1,v}}
\end{equation}

The model is trained with Adam optimizer \cite{adam}. 

\subsubsection{Other Implementation Details}

We add residual connections \cite{resnet} to positional attention layers and position-wise feed-forward networks. We add dropout \cite{dropout} to attention layers and feed-forward layers to reduce overfitting. 

\subsection{Discussion}

The main difference between our models and other self-attention-based sequential recommendation model like SASRec is that, while they relying on interaction between dot product attention mechanism and positional embedding to distinguish item in different positions, our models directly learn positional relations through a learnable matrix \(\mathbf{R}_l\). Using a single matrix to capture positional relations introduces \(n\times n\) additional parameters, where \(n\) represents the maximum length of the input sequence. However, it eliminates the need to compute \(\mathbf{Q}\) (query) and \(\mathbf{K}\) (key), thereby saving \(2d^2\) parameters required for the query and key projections.

FPARec differs from PARec by factorizing positional attention matrix \(\mathbf{R}_l\) into two low-rank matrices \(\mathbf{R}_{1,l}\) and \(\mathbf{R}_{2,l}\), introducing \(2kn\) additional parameters. Generally, \(k\) is significantly smaller than \(n\), thus \(2kn << n^2\), resulting in a substantially reduced parameter count. More detailed comparison on parameter count can be found in Table \ref{table:parameter_count}.

\begin{table*}[t]
    \vspace{-2em}
    \centering
	\caption{Comparison of parameter count in the attention layer between the proposed PARec and FPARec compared with SASRec}
	\label{table:parameter_count}
	\renewcommand{\arraystretch}{1.2}
	\begin{tabular}{lcccccccccccc}
	\hline
             & Number of parameters per layer    \\
    \hline\hline
    SASRec   & \(3d^2\)  \\
    PARec    & \(d^2 + n^2\)  \\
    FPARec   & \(d^2 + 2kn\) \\
\hline
	\end{tabular}
\vspace{-1em}
\end{table*}

\section{Experiments}

In this section, we conduct experiments on five common sequential recommendation datasets to validate the performance of our model. 

\subsection{Datasets and Evaluation Metrics}

\begin{table}[b]
\vspace{-2em}
\caption{Statistics of the datasets after preprocessing.}\label{table:dataset}
\centering
\renewcommand{\arraystretch}{1.2}
\begin{tabular}{l|r|r|r|r} 
\hline
\textbf{Dataset} & \textbf{\# Users} & \textbf{\# Items} & \textbf{\# Interactions} & \textbf{Avg. length} \\
\hline\hline
ML-1m & 6,040 & 3,416 & 1.00M & 165.50 \\
Beauty & 22,363 & 12,101 & 0.20M & 8.88 \\
Sports & 35,598 & 18,357 & 0.30M & 8.32 \\
Toys & 19,412 & 11,924 & 0.17M & 8.63 \\
Yelp & 30,431 & 20,033 & 0.32M & 10.40 \\
\hline
\end{tabular}
\vspace{-10pt}
\end{table}

The statistics of the experiment datasets are listed in Table \ref{table:dataset}. Following is a brief introduction for each dataset. 
\begin{itemize}
\item \textbf{MovieLens} \cite{movie_lens}: The MovieLens dataset was collected by an online movie recommendation platform. We use the version which contains one million rates (ML-1m).

\item \textbf{Beauty, Sports and Toys} \cite{amazon}: Three subcategories of amazon review datasets. We choose Beauty, Sports and outdoors, Toys and games to test our model. Each review is used as an interaction. 

\item \textbf{Yelp}\footnote{https://www.yelp.com/dataset}: The Yelp dataset consists of business, review, and user data. We use the transaction records after January 1, 2019, in the experiment. 
\end{itemize}
We follow the practice in \cite{sasrec} to preprocess the datasets. We use reviews or ratings as interactions. We sort the interactions in chronological order and group interactions by users to form training sequences. We split each user’s interaction sequence into three parts: the most recent interaction is used for testing, the second most recent interaction is used for validation, and all remaining interactions are used for training. We only keep items and users with more than five interactions. The statistics of datasets are shown in Table \ref{table:dataset}.

The performance of models is evaluated using Hit Rate@10 and NDCG@10\cite{sasrec}. HR@10 calculates the proportion of times that the ground-truth item is ranked among the top-10 items by the model. NDCG@10 takes into account the placement and assigns higher weight to higher placements, emphasizing the importance of ranking relevant items more prominently. Since some recent research suggests that sample metrics may not be consistent with their exact version\cite{sampled_metrics,sampling_strategies}, we rank the ground truth item against all other items in the dataset when evaluating the model performance. 

\subsection{Baseline Methods}

We choose three baselines to compare the performance with our model:

\begin{itemize}
\item \textbf{SASRec} \cite{sasrec} is state of the art method for sequential recommendation. It employs the standard scaled dot-product self-attention mechanism in a unidirectional fashion. 

\item \textbf{Linear Attention}. We replace the self-attention module in SASRec with linearized attention\cite{transformer_rnns}.

\item \textbf{Token Mixing}. The self-attention layer of SASRec is replaced by the token mixing layer proposed in \cite{mlp_mixer}. To make the result fair, we mask the future positional in token mixing layers. 
\end{itemize}

For SASRec, we use the code provided by the author. Other methods were implemented in Tensorflow. We consider maximum sequence length in the range \{50, 200, 500\} for ML-1m and in the range \{50, 100, 200\} for other datasets. Other hyperparameters are either following author's suggestion or tuned on the validation set. We train each model 3 times and report the median performance on the test set.

\begin{table*}[t]
    \vspace{-1em}
    \centering
	\caption{Performance comparison of different methods on five datasets. The best performance on each dataset is boldfaced, and the second-best performance is underlined.}
	\label{table:performance}
	\renewcommand{\arraystretch}{1.2}
	\begin{tabular}{llccccccccccccc}
	\hline
		Datasets & Metric & SASRec & Linear Attention & Token Mixing & PARec & FPARec  \\
	\hline\hline
\multirow{2} * {ML-1m}
 &HR@10   & \underline{0.2346} & 0.2235 & 0.2296 & 0.2255 & \textbf{0.2419} \\
 &NDCG@10 & \underline{0.1186} & 0.1089 & 0.1149 & 0.1130 & \textbf{0.1232} \\
\hline
\multirow{2} * {Beauty}
 &HR@10   & \underline{0.0813} & 0.0785 & 0.0798 & 0.0806 & \textbf{0.0821} \\
 &NDCG@10 & \textbf{0.0405} & 0.0379 & 0.0398 & 0.0395 & \underline{0.0402} \\
\hline
\multirow{2} * {Sports}
 &HR@10   & 0.0460 & 0.0462 & 0.0458 & \underline{0.0470} & \textbf{0.0478} \\
 &NDCG@10 & \underline{0.0233} & 0.0228 & \textbf{0.0235} & 0.0230 & \textbf{0.0235} \\
\hline
\multirow{2} * {Toys}
 &HR@10   & 0.0843 & 0.0847 & 0.0852 & \underline{0.0859} & \textbf{0.0861} \\
 &NDCG@10 & 0.0410 & 0.0415 & \underline{0.0416} & \underline{0.0416} & \textbf{0.0421} \\
\hline
\multirow{2} * {Yelp}
 &HR@10   & 0.0643 & 0.0591 & 0.0632 & \underline{0.0647} & \textbf{0.0654} \\
 &NDCG@10 & \underline{0.0369} & 0.0350 & 0.0354 & 0.0368 & \textbf{0.0372} \\
\hline
	\end{tabular}
\vspace{-1em}
\end{table*}

\subsection{Performance Comparison}

In the domain of sequential recommendation tasks, our proposed method FPARec, utilizing factorized attention layers, showcases outstanding performance, topping the charts in HR@10 across all five datasets as illustrated in Table \ref{table:performance}. PARec, another variant we introduced that utilizes positional attention layers, also performs admirably, achieving the second-best results in HR@10 for three out of the five datasets. In terms of NDCG@10, FPARec continues to excel, outperforming all competing methods in four of the datasets, highlighting its robustness and effectiveness. The comparative results distinctly demonstrate the efficiency of our method in handling sequential recommendation challenges. Notably, FPARec’s performance surpasses that of PARec, indicating the benefits of factorizing a positional attention matrix into two low-rank matrices. This factorization leads to a reduction in model parameters and enhances data efficiency during training, contributing to the overall superior performance of FPARec.

\subsection{Ablation Study}

Since we can handcraft some attention pattern to mimic the effect of putting more attention weight on recent item, it is interesting to know whether our learned positional embedding is better than the carefully designed fixed attention patterns. To verify the effectiveness of our proposed positional attention mechanism, we replace the positional attention layers in our model with several handcrafted attention patterns below:

\begin{itemize}
\item \textbf{Average}: This attention pattern means that all previous items will receive equal weight for predicting the next item.

\item \textbf{Linear Annealing}: In this attention mechanism, recent items receive a higher weight. Earlier items receive a lower weight. 

\item \textbf{Exponential Annealing}: In this attention mechanism, attention weights decrease exponentially for earlier items, meaning that more recent items have substantially higher weights compared to older items.

\end{itemize}


The experiment results are shown in Table \ref{table:ablation}. We observed that our proposed model, FPARec, consistently outperforms the fixed attention patterns in terms of HR@10 on both the ML-1m and Yelp datasets. This underscores the effectiveness of FPARec in adapting to the diverse patterns of user interactions, especially in datasets with extensive user interactions like Yelp. 

Interestingly, while FPARec shows overall superiority, the Exponential Annealing pattern slightly surpasses FPARec in NDCG@10 on the ML-1m dataset. This exception highlights the adaptability of Exponential Annealing in scenarios with a smaller user base, where recent interactions play a more pivotal role. However, it is important to note that this does not detract from the overall performance and versatility of FPARec, which delivers robust results across different metrics and datasets. On the Yelp dataset, FPARec not only maintains its lead in HR@10 but also excels in NDCG@10, reinforcing its applicability and effectiveness in handling larger, more complex datasets. 

In conclusion, our findings suggest that while simpler fixed attention patterns like Exponential Annealing may have their niche advantages, FPARec stands out as a more versatile and powerful tool for sequential recommendation tasks, adapting effectively to both small and large datasets.

\begin{table}[t]
    \centering
    \vspace{-2em}
	\caption{Ablation study of our methods on ML-1m and Yelp dataset}
	\label{table:ablation}
	{
		\begin{tabular}{l|cc|cc}
			\hline
			 &  \multicolumn{2}{c|}{ML-1m} &
			\multicolumn{2}{c}{Yelp} \\
			\hline
		     & HR@10 & NDCG@10 & HR@10 &NDCG@10 \\
		    \hline\hline
			FPARec         & \textbf{0.2419} & \textbf{0.1232} & \textbf{0.0654} & \textbf{0.0372}  \\
			\hline
			PARec          & 0.2255 & 0.1130 & 0.0647 & 0.0368   \\
			SASRec      & 0.2346 & 0.1186 & 0.0643 & 0.0369  \\
			\hline
			+Average    & 0.2025 & 0.0986 & 0.0560 & 0.0336 \\
			+Linear Annealing       & 0.2056 & 0.0997 & 0.0567 & 0.0338 \\
 			+Exponential Annealing  & 0.2386 & \textbf{0.1283} & 0.0574 & 0.0337 \\
 			\hline
		\end{tabular}
	}
    \vspace{-1em}
\end{table}

\begin{table*}[b]
    \vspace{-2em}
    \centering
	\caption{Impact of \(k\) in FPARec on ML-1m dataset. The maximum input sequence length is fixed at 500. }
	\label{table:influence_k}
	\renewcommand{\arraystretch}{1.2}
	\begin{tabular}{lcccccccccccccc}
	\hline
    k       & 10     & 20     & 30     & 40     & 50     & 60 \\
    \hline\hline
    HR@10   & 0.2368 & 0.2397 & 0.2404 & \textbf{0.2419} & 0.2389 & 0.2412 \\
    NDCG@10 & 0.1212 & 0.1220 & 0.1209 & 0.1232 & 0.1212 & \textbf{0.1243} \\
\hline
\vspace{-2em}
	\end{tabular}
\end{table*}

\begin{table*}[t]
    \vspace{-2em}
    \centering
	\caption{Impact of number of blocks in FPARec on ML-1m dataset.}
	\label{table:influence_num_blocks}
	\renewcommand{\arraystretch}{1.2}
	\begin{tabular}{lcccccccccccc}
	\hline
    number of blocks       & 1     & 2     & 3 \\
    \hline\hline
    HR@10   & 0.2275 & 0.2419 & \textbf{0.2500} \\
    NDCG@10 & 0.1126 & 0.1232 & \textbf{0.1303} \\
\hline
	\end{tabular}
\vspace{-1em}
\end{table*}

\subsection{Impact of Factorizing Dimension \(k\)}

FPARec introduces a hyperparameter: the factorizing dimension, \(k\), of the factorized positional attention layers. To investigate the influence of \(k\), we fix the maximum sequence length at 500 and conduct experiments with various values of \(k\) on the ML-1m dataset. The results of these experiments are presented in Table \ref{table:influence_k}. As can be seen from the table, the model demonstrates a relative insensitivity to the choice of \(k\). Performance remains robust, providing \(k\) is sufficiently large, typically at or above 20.

\subsection{Impact of number of blocks}

We alter the number of attention blocks to inspect how the number of blocks influences model performance. The results, presented in Table \ref{table:influence_num_blocks}, show our model benefits from stacking more attention blocks. The underlying reason for this improvement is that increasing number of blocks allows the model to capture more complex patterns and dependencies in the data. With a single attention block, the model's capacity is limited, and it may not be able to learn and represent the intricate relationships present in the dataset. 

As the number of blocks is increased to two, there is a noticeable improvement in both HR@10 and NDCG@10, indicating that the model is able to provide more accurate and relevant recommendations. This trend continues with the addition of a third attention block, leading to highest observed performance in terms of both metrics. 

It is important to note, however, that while adding more attention blocks can enhance the model's capability to understand and represent the data, it also increases the computational complexity and the number of parameters in the model. This can lead to longer training times and require more memory and computational resources. As a result, we use two attention blocks in our model. 

\subsection{Visualization of Learned Attention}

\begin{figure}[t]
    \vspace{-1em}
    \centering
    \begin{subfigure}[b]{0.23\textwidth}
        \centering
        \includegraphics[width=\textwidth]{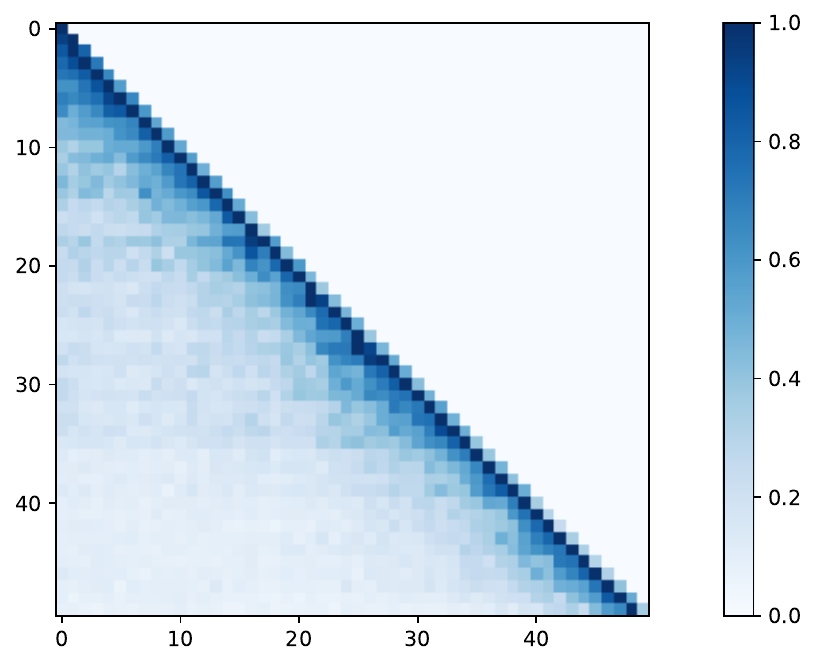}
        \caption{PARec-1, 50}
        \label{fig:pa_50}
    \end{subfigure}
    \hfill
    \begin{subfigure}[b]{0.23\textwidth}
        \centering
        \includegraphics[width=\textwidth]{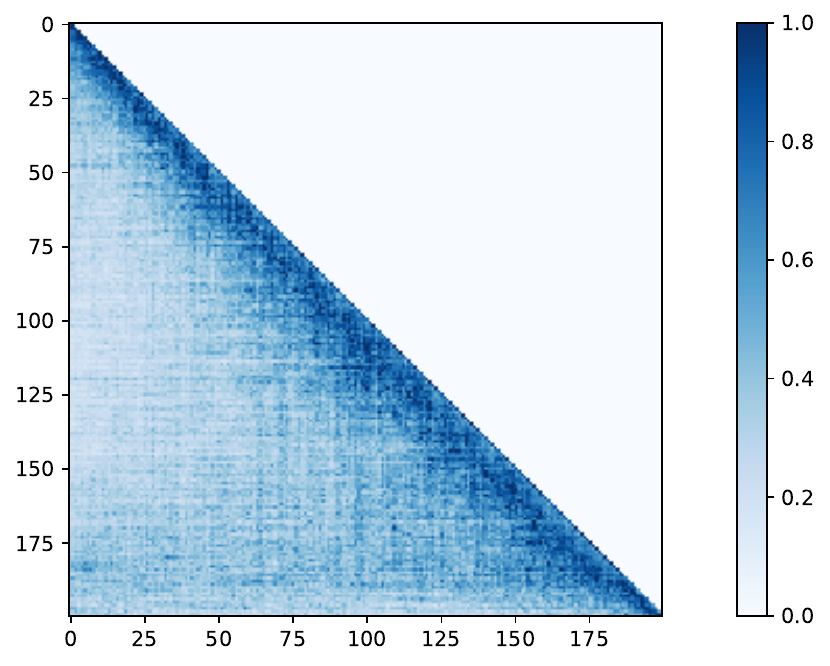}
        \caption{PARec-1, 200}
        \label{fig:pa_200}
    \end{subfigure}
    \hfill
    \begin{subfigure}[b]{0.23\textwidth}
        \centering
        \includegraphics[width=\textwidth]{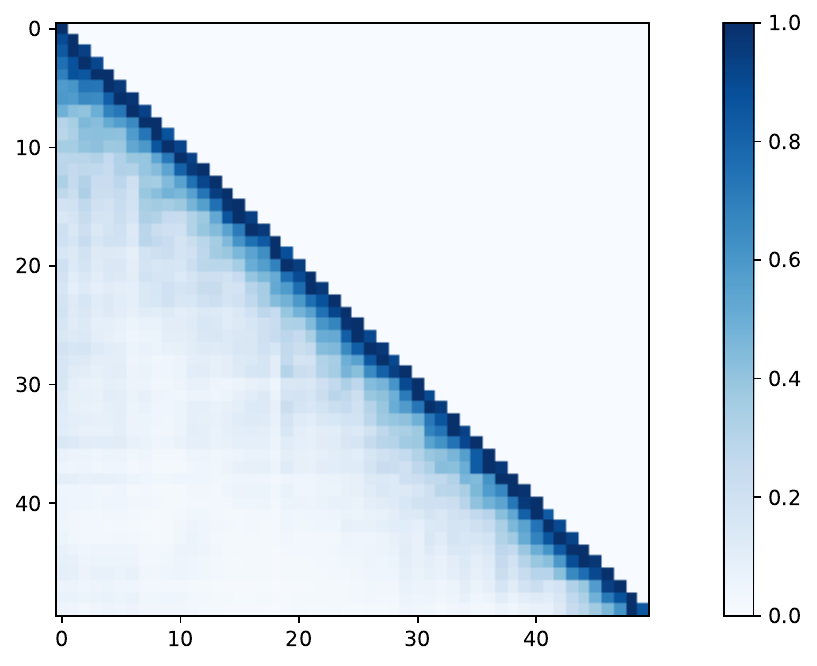}
        \caption{FPARec-1, 50}
        \label{fig:fpa_50}
    \end{subfigure}
    \hfill
    \begin{subfigure}[b]{0.23\textwidth}
        \centering
        \includegraphics[width=\textwidth]{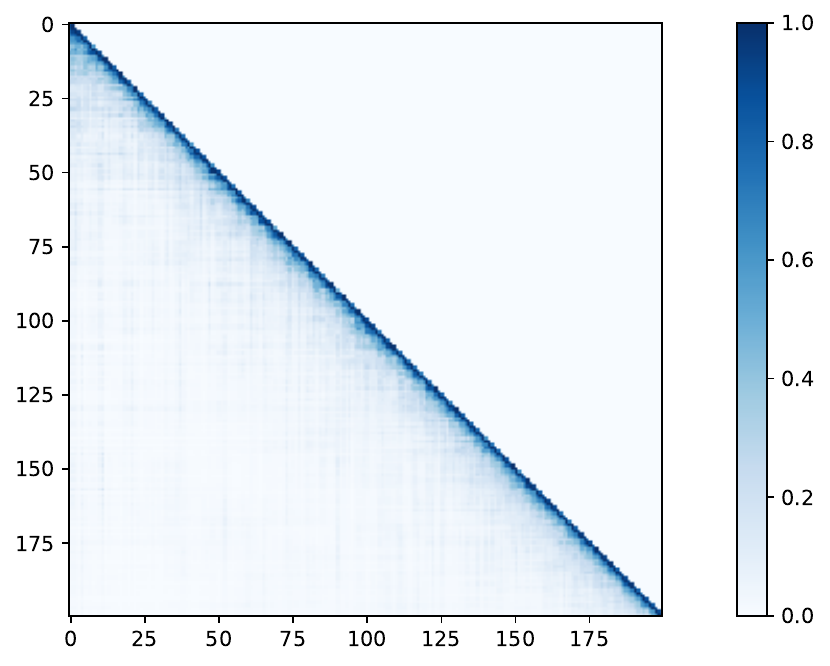}
        \caption{FPARec-1, 200}
        \label{fig:fpa_200}
    \end{subfigure}
    
    \begin{subfigure}[b]{0.23\textwidth}
        \centering
        \includegraphics[width=\textwidth]{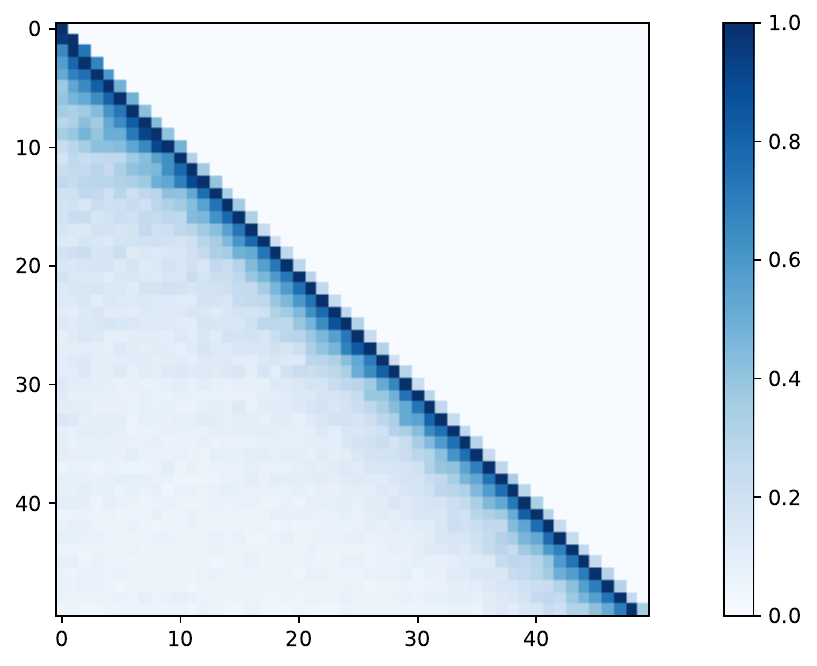}
        \caption{PARec-2, 50}
        \label{fig:pa_50_2}
    \end{subfigure}
    \hfill
    \begin{subfigure}[b]{0.23\textwidth}
        \centering
        \includegraphics[width=\textwidth]{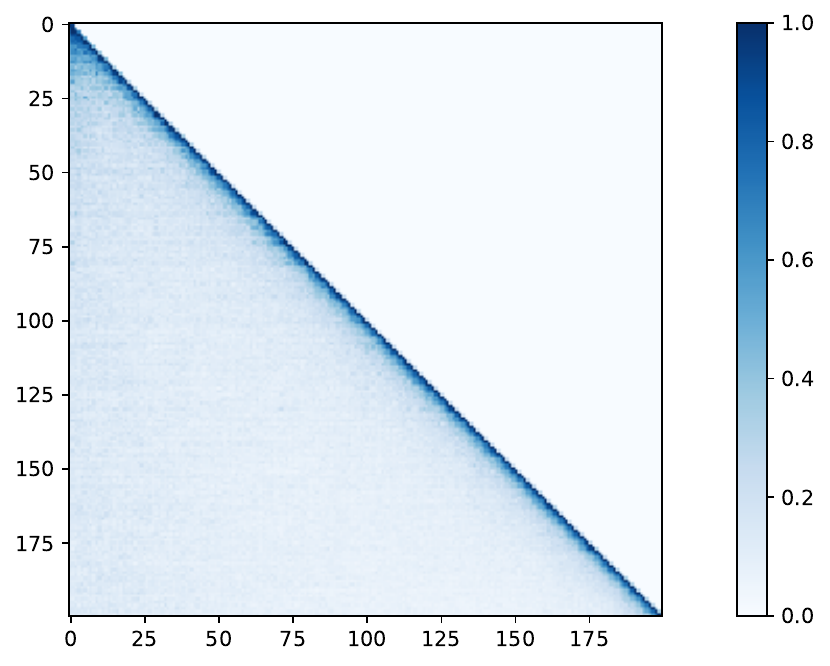}
        \caption{PARec-2, 200}
        \label{fig:pa_200_2}
    \end{subfigure}
    \hfill
    \begin{subfigure}[b]{0.23\textwidth}
        \centering
        \includegraphics[width=\textwidth]{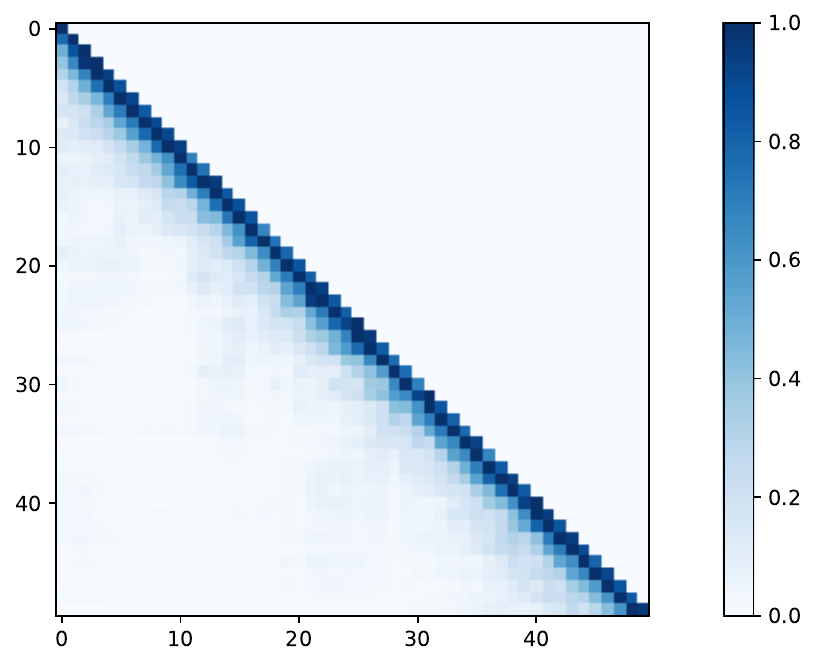}
        \caption{FPARec-2, 50}
        \label{fig:fpa_50_2}
    \end{subfigure}
    \hfill
    \begin{subfigure}[b]{0.23\textwidth}
        \centering
        \includegraphics[width=\textwidth]{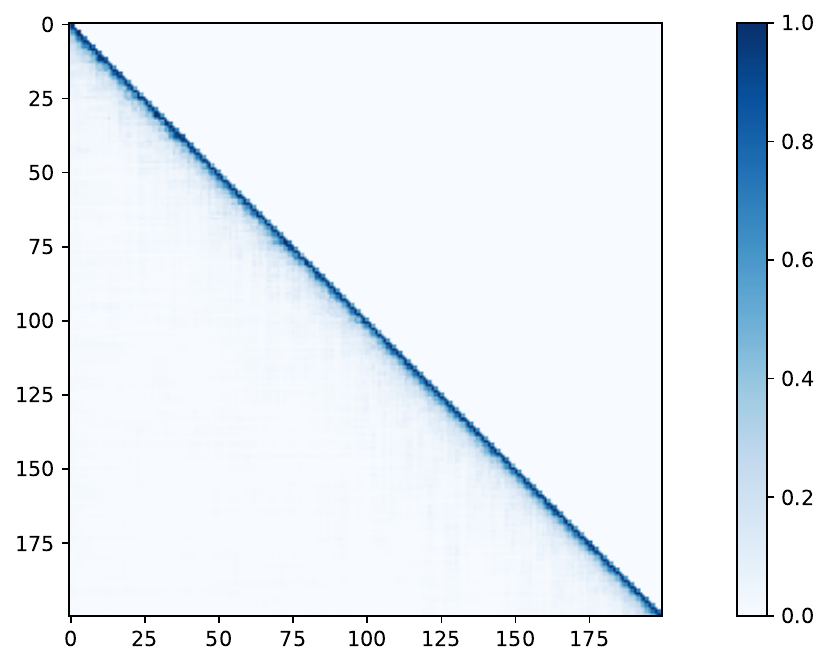}
        \caption{FPARec-2, 200}
        \label{fig:fpa_200_2}
    \end{subfigure}
    
    \caption{Visualization of learned attention pattern. Each row is normalized by the maximum value in that row for better comparison. The first row shows the visualization of the learned attention weight of the first attention layer of the model. The learned attention weight of the second attention layer is shown in the second row. Number 50 and 200 denotes the maximum input sequence length of the model.}
    \label{fig:learned_attention}
    \vspace{-2em}
\end{figure}

We train our model on the ML-1m dataset, setting the maximum input sequence length at either 50 or 200, and visualized the learned attention in the attention blocks, as shown in Fig \ref{fig:learned_attention}. Within each model configuration, the first attention layer displayed a more distributed weight pattern, whereas the second layer manifested more concentrated weights. This pattern suggests a hierarchical data representation inherent to the attention mechanism. Initially, the first attention layer offers a broader perspective across the sequence, capturing diverse contextual relationships from both recent and earlier items.

Conversely, the attention in the second layer becomes notably sharper, majorly emphasizing recent interactions. This resonates with the prevalent understanding in sequential recommendation tasks, wherein recent activities often bear more weight. Such a hierarchical attention setup allows the model to detect patterns over extended durations as well as more recent ones, leading to improved prediction accuracy.

The first layer's learned attention weight for PARec, when set to a maximum input sequence length of 200 as seen in Fig \ref{fig:pa_200}, appears more erratic compared to its counterpart for FPARec depicted in Fig \ref{fig:fpa_200}. Modeling extended sequences means the positional attention layer demands more parameters compared to its factorized variant, making it more susceptible to overfitting. The test performance of FPARec with an input sequence length capped at 200 yielded an HR@10 score of 0.2351. In comparison, PARec, under the same conditions, scored 0.2243. The superior results achieved by FPARec, given the same sequence length, align with its less chaotic visualization outcome.

\section{Relate Work}
Early works \cite{fpmc} in sequence recommendation used Markov chains to capture sequence patterns from user historical interactions. With the rapid development of neural networks, many methods started using CNN-based \cite{caser} and RNN-based \cite{gru4rec,gru4rec++} models for sequential recommendation and achieved good results. In recent years, many works have explored the use of attention mechanisms to improve the performance of sequence recommendations. SASRec\cite{sasrec} uses self-attention mechanism to model the user sequence information. BERT4Rec\cite{BERT4Rec} uses a bidirectional deep sequence model based on self-attention to make sequence recommendations. They all achieved outstanding results. The self-attention mechanism relies on positional encoding to model order of items in sequence. 

\section{Conclusion}

In this work, we explored the characteristics of positional embeddings in self-attention-based sequential recommendation models and showed that these embeddings tend to capture the relative distances between tokens. To leverage this finding, we introduced PARec and FPARec, two novel methods featuring learnable positional attention mechanisms. PARec utilizes a single matrix to learn positional relationships, while FPARec employs a factorized approach, reducing the number of parameters. Extensive experiments across multiple real-world datasets confirmed the efficacy of both PARec and FPARec in sequential recommendation tasks.

%
%
%
\bibliographystyle{splncs04}
\bibliography{cas-refs}

\newpage
\appendix
\section{Details on Drawing Fig \ref{fig:correlation}}

We give details on generating Figure \ref{fig:correlation} in this section. We trained the SASRec model on the ML-1m dataset to convergence, using maximum input sequence lengths of 50 and 200. Post training, we extracted the positional embeddings from the model. To calculate the correlation between positional embeddings at different positions, we used the expression \(\exp{(\frac{P_iP_j^T}{\sqrt{d}})}\), which yields an \(N \times N\) matrix. For better visualization, each row of the matrix was divided by its maximum value, ensuring that the highest attention score in each row equates to 1. The pseudocode detailing the steps to generate this figure is provided below.

\begin{figure}
\begin{lstlisting}
"""Code to process postional embeddings and draw the figure"""
correlation = np.dot(positional_embeddings, positional_embeddings.T) 
correlation /= np.sqrt(positional_embeddings.shape[-1])
correlation = np.exp(correlation)
correlation = np.tril(correlation) # Mask future positions
correlation = correlation / np.max(correlation, axis=-1)
plt.imshow(correlation, cmap='Blues')
\end{lstlisting}
\end{figure}

\section{Visualization of the Handcrafted Attention Patterns Discussed in the Ablation Study}

\begin{figure}
\captionsetup{font=scriptsize}
    \centering
    \begin{subfigure}[b]{0.23\textwidth}
        \centering
        \includegraphics[width=\textwidth]{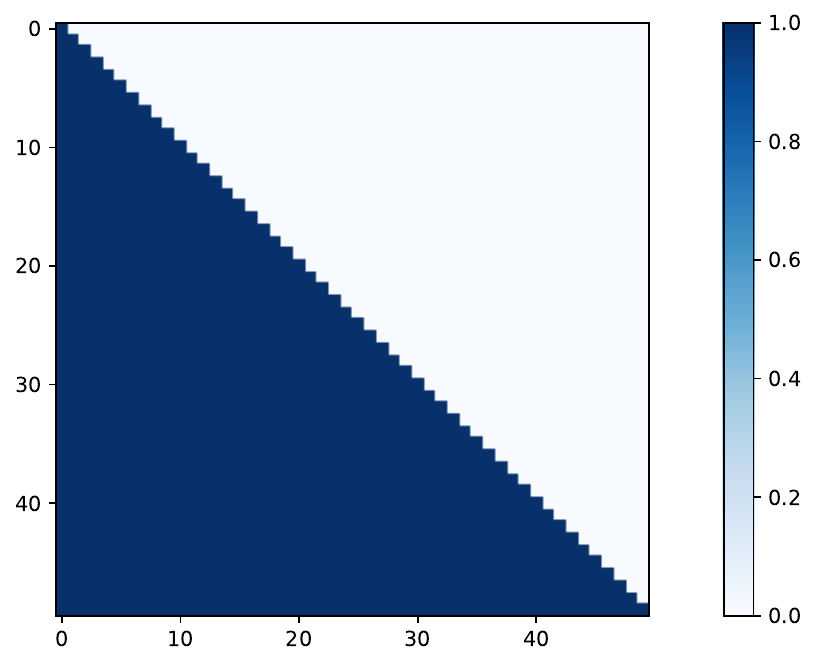}
        \caption{Average, 50}
        \label{fig:average50}
    \end{subfigure}
    \hfill
    \begin{subfigure}[b]{0.23\textwidth}
        \centering
        \includegraphics[width=\textwidth]{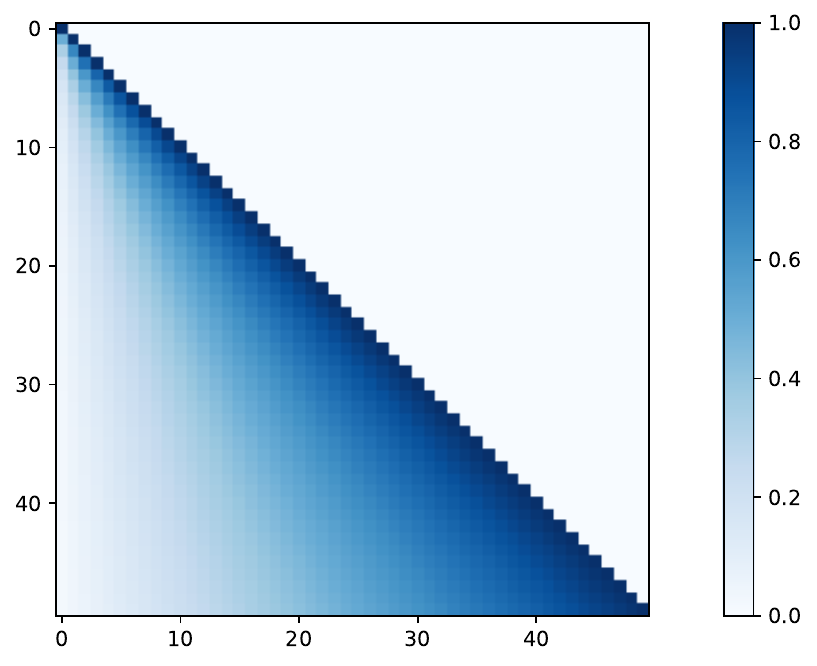}
        \caption{Linear, 50}
        \label{fig:linear50}
    \end{subfigure}
    \hfill
    \begin{subfigure}[b]{0.23\textwidth}
        \centering
        \includegraphics[width=\textwidth]{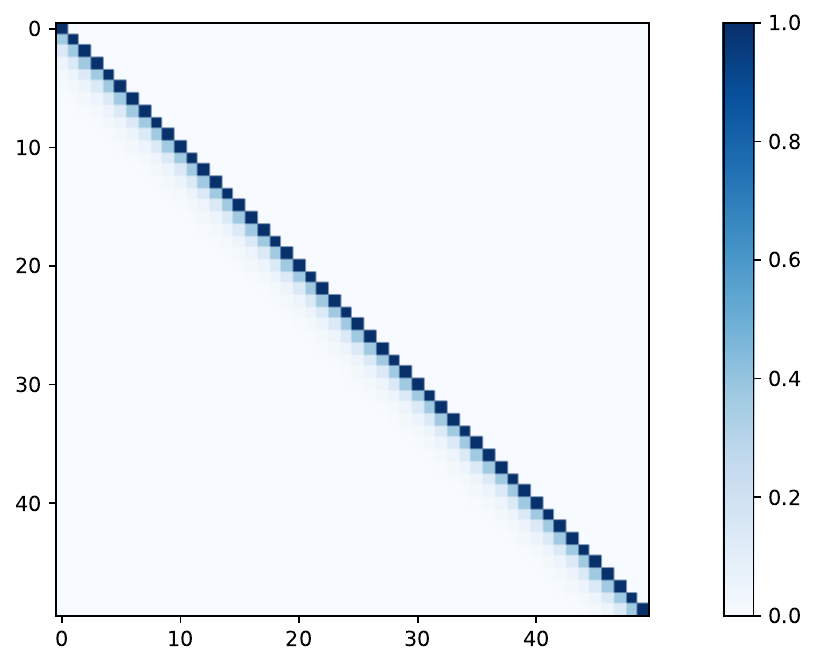}
        \caption{Exponential, 50}
        \label{fig:exponential50}
    \end{subfigure}
    
    \begin{subfigure}[b]{0.23\textwidth}
        \centering
        \includegraphics[width=\textwidth]{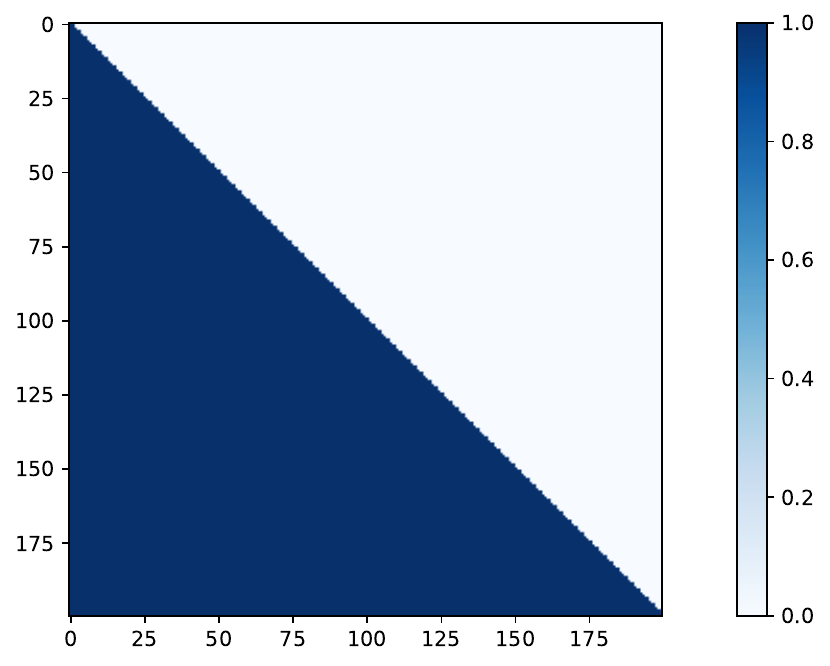}
        \caption{Average, 200}
        \label{fig:average200}
    \end{subfigure}
    \hfill
    \begin{subfigure}[b]{0.23\textwidth}
        \centering
        \includegraphics[width=\textwidth]{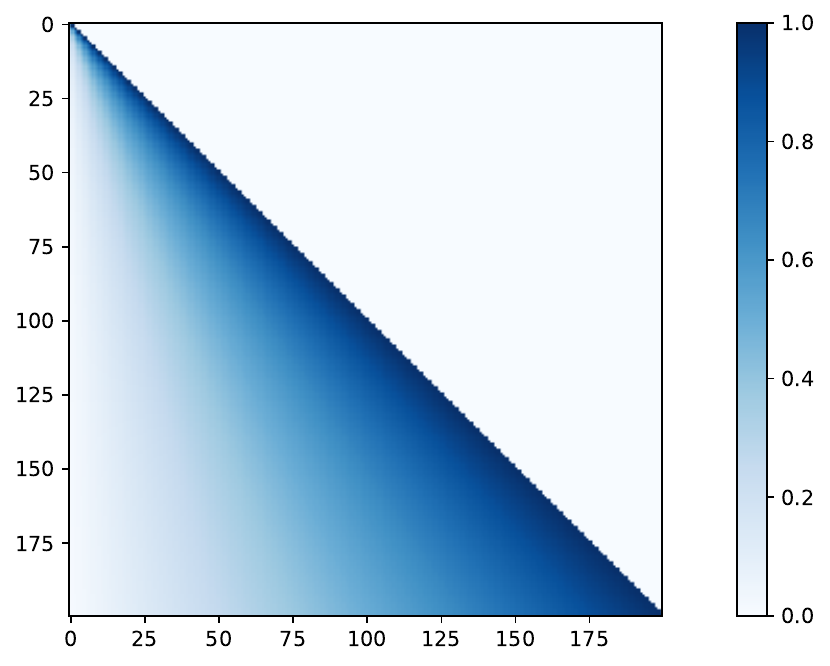}
        \caption{Linear, 200}
        \label{fig:linear200}
    \end{subfigure}
    \hfill
    \begin{subfigure}[b]{0.23\textwidth}
        \centering
        \includegraphics[width=\textwidth]{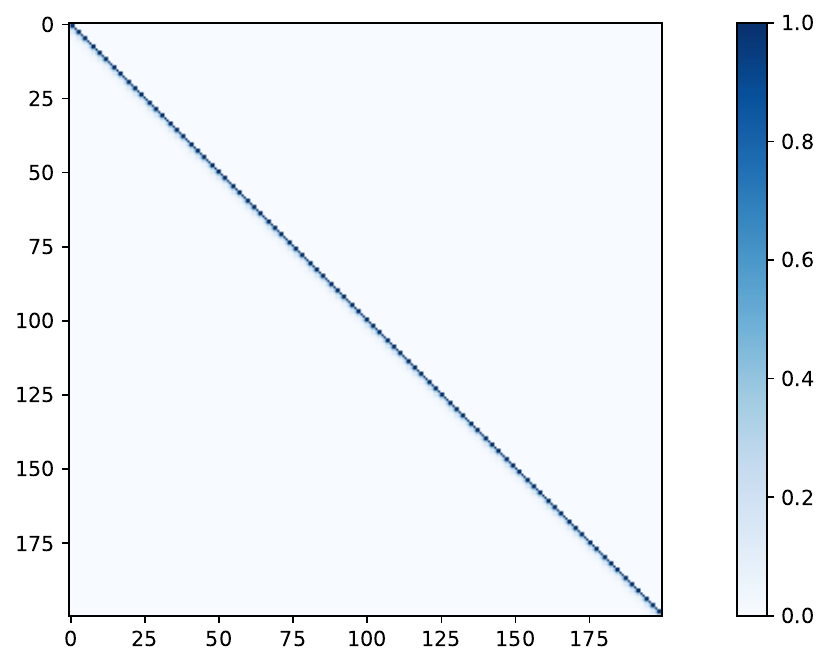}
        \caption{Exponential, 200}
        \label{fig:exponential200}
    \end{subfigure}

    \captionsetup{font=small}
    \caption{Visualization of predefined attention patterns, with \textbf{Average}, \textbf{Linear Annealing} and \textbf{Exponential Annealing} representing different strategies. For better comparison, each row has been normalized by its maximum value. The input sequence lengths are indicated by 50 and 200. Detail may require zooming in to view.}

    \label{fig:handcrafted_attention}
    \vspace{-10pt}
\end{figure}

We visualize the handcrafted attention patterns discussed in ablation study. The result is in Figure \ref{fig:handcrafted_attention}. We use \(a_{ij}\) to denote the fixed attention weight for position \(i\) when attending to position \(j\). We provide the mathematical definitions for these handcraft attention patterns as follows:

\textbf{Average}:
\begin{equation}
a_{ij}^{avg} =
  \begin{cases}
    1       & \quad \text{if } i >= j \\
    0       & \quad \text{otherwise}
  \end{cases}
\end{equation}

\textbf{Linear Annealing}:
\begin{equation}
a_{ij}^{lin} =
  \begin{cases}
    j       & \quad \text{if } i >= j \\
    0       & \quad \text{otherwise}
  \end{cases}
\end{equation}

\textbf{Exponential Annealing}:
\begin{equation}
a_{ij}^{exp} =
  \begin{cases}
    e^{j-i}       & \quad \text{if } i >= j \\
    0       & \quad \text{otherwise}
  \end{cases}
\end{equation}

Note \(a_{ij}^{avg}\), \(a_{ij}^{lin}\) and \(a_{ij}^{exp}\) are unnormalized attention scores. The normalized attention scores should be 

\begin{equation}
\tilde{a}_{ij} = \frac{a_{ij}}{\sum_{k=1}^{n}a_{ik}}
\end{equation}

\(n\) is the maximum length of input sequences for the model. 

\end{document}